\documentclass[longauth]{aa} 
\usepackage{natbib,twoopt}
\bibpunct{(}{)}{;}{a}{}{,} 

\usepackage{graphicx}
\usepackage{txfonts}
\begin{document}

	\title{ Molecular gas in AzTEC/C159: a star-forming disk galaxy 
		1.3\,Gyr after the Big Bang}

   \author{E.F. Jim\'enez-Andrade
	\inst{\ref{aifa},\ref{imprs}}
	\and
	B.~Magnelli\inst{\ref{aifa}}
	\and
	A.~Karim\inst{\ref{aifa}}
	\and
	G.~C.~Jones\inst{\ref{nmt}, \ref{nrao}}
	\and
	C.~L.~Carilli\inst{\ref{nrao}, \ref{cambridge}}
	\and
	E.~Romano-D\'iaz\inst{\ref{aifa}}
	\and 
	C.~G\'omez-Guijarro\inst{\ref{dark}}
	\and
	S.~Toft\inst{\ref{dark}}
	\and 
	F.~Bertoldi\inst{\ref{aifa}}
	\and
	D.~A.~Riechers\inst{\ref{cornell}}
	\and
	E.~Schinnerer\inst{\ref{hd}}
	\and 
	M.~Sargent\inst{\ref{sussex}}
	\and 
	M.~J.~Micha{\l}owski\inst{\ref{poland}}
	\and
	F.~Fraternali\inst{\ref{bologna}, \ref{groningen}}
	\and 
	J.~G.~Staguhn\inst{\ref{goddard}, \ref{hopkins}}
	\and 
	V.~Smol{\v c}i\'c\inst{\ref{zagreb}}
	\and
	M.~Aravena\inst{\ref{dp}}
	\and
	K.~C.~Harrington\inst{\ref{aifa},\ref{imprs}}
	\and 
	K.~Sheth\inst{\ref{nasa}}
	\and 	
	P.~L.~Capak\inst{\ref{caltech1},\ref{caltech2}}
	\and
	A.~M.~Koekemoer\inst{\ref{stsi}}
	\and 
	E.~van Kampen\inst{\ref{eso}}
	\and 
	M.~Swinbank\inst{\ref{durham}}
	\and 
	A.~Zirm\inst{\ref{dark},\ref{software}}
	\and
	G.~E.~Magdis\inst{\ref{dark},\ref{greece}}
	\and
	F.~Navarrete\inst{\ref{mpifr}}
}
	
	\institute{Argelander Institute for Astronomy, University of Bonn, Auf dem H\"ugel 71, D-53121 Bonn, Germany\label{aifa}
		\and
		International Max Planck Research School of Astronomy and Astrophysics at the Universities of Bonn and Cologne\label{imprs}
		\and 
		Physics Department, New Mexico Institute of Mining and Technology, 801 Leroy Pl, Socorro, NM 87801, USA\label{nmt}
		\and 
		National Radio Astronomy Observatory, 1003 Lopezville Road, Socorro, NM 87801, USA\label{nrao}
		\and 
		Cavendish Astrophysics Group, University of Cambridge, Cambridge, CB3 0HE, UK\label{cambridge}
		\and 
		Dark Cosmology Centre, Niels Bohr Institute, University of Copenhagen, Juliane Maries Vej 30, DK-2100 Copenhagen, Denmark\label{dark}
		\and
		Department of Astronomy, Cornell University, Space Sciences Building, Ithaca, NY 14853, USA\label{cornell}
		\and 
		Max Planck Institute for Astronomy, K\"onigstuhl 17, 69117 Heidelberg, Germany\label{hd}
		\and 
		Astronomy Centre, Department of Physics and Astronomy, University of Sussex, Brighton, BN1 9QH, UK\label{sussex}
		\and 
		Astronomical Observatory Institute, Faculty of Physics, Adam
		Mickiewicz University, ul.~S{\l}oneczna 36, 60-286 Pozna{\'n}, Poland\label{poland}
		\and
		Department of Physics and Astronomy, University of Bologna, viale Berti Pichat 6/2, I-40127 Bologna, Italy\label{bologna}
		\and
		Kapteyn Astronomical Institute, Postbus 800, NL-9700 AV Groningen, the Netherlands\label{groningen}
		\and 		
		NASA Goddard Space Flight Center, Code 665, Greenbelt, MD 20771, USA\label{goddard}
		\and
		Department of Physics and Astronomy, Johns Hopkins University, Baltimore, MD 21218, USA\label{hopkins}
		\and 
		Department of Physics, Faculty of Science, University of Zagreb, Bijeni{\v c}ka cesta 32, 10000 Zagreb, Croatia\label{zagreb}
		\and
	    N\'ucleo de Astronom\'ia, Facultad de Ingenier\'ia y Ciencias, Universidad Diego Portales, Av. Ej\'ercito 441, Santiago, Chile\label{dp}
	    \and 
	    Science Mission Directorate, NASA Headquarters, Washington, DC 20546-0001, USA\label{nasa}
	    \and 
		Infrared Processing and Analysis Center, California Institute of Technology, MC 100-22, 770 South Wilson Ave., Pasadena, CA 91125, USA\label{caltech1}
		\and
		Spitzer Science Center, California Institute of Technology, Pasadena, CA 91125, USA\label{caltech2}
		\and
		 Space Telescope Science Institute, 3700 San Martin Drive,
		Baltimore, MD 21218, USA\label{stsi}
		\and 
		European Southern Observatory
		Karl-Schwarzschild-Strasse 2
		D-85748 Garching bei Muenchen, Germany\label{eso}
		\and 
		Centre for Extragalactic Astronomy, Durham University, South Road, Durham DH1 3LE, UK\label{durham}
		\and 
		Greenhouse Software, 3rd Floor, 110 5th Avenue, New York, NY 10011, USA\label{software}
		\and 
		Institute for Astronomy, Astrophysics, Space Applications and Remote Sensing, National Observatory of Athens, GR-15236 Athens, Greece\label{greece}
		\and
		 Max Planck Institute for Radioastronomy, Auf dem H\"ugel 69, D-53121 Bonn, Germany\label{mpifr}\\
		\email{ericja@astro.uni-bonn.de}
	}
	
	\date{Received October 2017; accepted February 2018}
	
	
	\abstract
	{
	We studied the molecular gas properties of AzTEC/C159, a star-forming disk galaxy at $z=4.567$, in order to better constrain the nature of the high-redshift end of the sub-mm selected galaxy (SMG) population. 
	We secured $^{12}$CO molecular line detections for the \emph{J}=2$\to$1 and \emph{J}=5$\to$4  transitions using the Karl G. Jansky Very Large Array (VLA) and the NOrthern Extended Millimeter Array (NOEMA) interferometer. 
	The broad ({\sc FWHM}$\sim750\,{\rm km\,s}^{-1}$) and tentative  double-peaked 	profiles of both $^{12}$CO lines are consistent with an extended molecular gas reservoir, which is distributed in a rotating disk as previously revealed from [CII]~158\,$\mu$m line observations.
	 Based on the $^{12}$CO(2$\to$1) emission line we derived  $L'_{\rm{CO}}=(3.4\pm0.6)\times10^{10}{\rm \,K\,km\,s}^{-1}{\rm \,pc}^{2}$, that yields  a molecular gas mass of $M_{\rm H_2 }(\alpha_{\rm CO}/4.3)=(1.5\pm0.3)\times 10^{11} \,$M$_\odot$ and unveils a gas-rich system with $\mu_{\rm gas}(\alpha_{\rm CO}/4.3)\equiv M_{\rm H_2}/M_\star=3.3\pm0.7$.  The extreme star formation efficiency (SFE) of AzTEC/C159, parametrized by the ratio $L_{\rm{IR}}/L'_{\rm{CO}}=(216\pm80)\, {\rm L}_{\odot}{\rm \,(K\,km\,s}^{-1}{\rm \,pc}^{2})^{-1}$,  is comparable to merger-driven starbursts such as local ultra-luminous infrared galaxies (ULIRGs) and SMGs.  Likewise, 
 the $^{12}$CO(5$\to$4)/CO(2$\to$1) line brightness temperature ratio of $r_{52}= 0.55\pm 0.15$ is consistent with high excitation conditions, similar to that observed in SMGs.
 Based on  mass budget considerations we   constrained the value for the $L'_{\text{CO}}$~--~H$_2$ mass conversion factor in AzTEC/C159, i.e.  \hbox{$\alpha_{\text{CO}}=3.9^{+2.7}_{-1.3}{\rm \,M}_{\odot}{\rm \,K}^{-1}{\rm \,km}^{-1}{\rm \,s\,pc}^{-2}$}, that is consistent with a self-gravitating molecular gas distribution as observed in local star-forming disk galaxies. 
 Cold gas streams from cosmological filaments might be fueling a gravitationally unstable gas-rich disk in AzTEC/C159, which breaks into giant clumps forming stars as efficiently as in merger-driven systems and generate high gas excitation. These results support the evolutionary connection between  AzTEC/C159-like  systems and massive quiescent disk galaxies at $z\sim2$. 
 }
	\keywords{galaxies: high-redshift -- galaxies: formation -- galaxies: ISM -- ISM: molecules
	}

 	\titlerunning{Molecular gas in AzTEC/C159}
	\authorrunning{E.F. Jim\'enez-Andrade et al. }
	\maketitle
%

\section{Introduction}
Sub-mm selected galaxies (SMGs), gas-rich starbursts at high redshifts  \citep[e.g.,][]{blain02,tacconi06,tacconi08, narayanan10, casey14}, might be the progenitors of compact quiescent galaxies ($\log(M_\star/ M_\odot)>11$) at $z\sim2$  \citep[e.g.,][]{simpson14, toft14,ikarashi15, oteo16, oteo17}.
It is believed that the intense starburst episode can be followed by an   active galactic nucleus (AGN) phase that eventually quenches the star formation \citep{hopkins06, wuyts10, hickox12, steinhardtspeagle14}. 
However, there is a lack of consensus on  the physical mechanisms driving the extreme production of stars in  SMGs. It has been proposed that  compact starbursts might be fueled via major gas-rich mergers \citep[e.g., ][]{walter09, narayanan10, hayward11, alaghband12, hayward12, riechers13, riechers14}. An alternative scenario involves the  smooth infall and accretion of cold gas from the intergalactic medium that could also drive intense star formation in massive highly unstable high-redshift galaxies
 \citep[e.g., ][]{dekel09, keres09a, keres09b, dave10, hodge12, romano-diaz14, feng15, anglesalcazar17}. \\

To constrain the origin of SMGs, and their possible evolutionary path, much effort has been put into building statistically complete and unbiased samples of these objects \citep[e.g.,][]{banerji11, yun12, hodge13, strandet16, brisbin17, michalowski17}. {\rm Particular emphasis has been placed on the recently discovered high-redshift tail ($4<z<6$) of the SMG population    \citep[e.g.,][]{daddi09b, daddi09, capak08, capak11, coppin09, knudsen10, smolcic11, barger12, walter12, ivison16}. By exploring the physical properties of these  $z>4$ systems one might  further strengthen the evolutionary link between high-redshift SMGs and massive quiescent galaxies at $z\sim2$ \citep[e.g.,][]{fudamoto17a}, and provide  constraints on cosmological models that aim to reproduce the extreme and massive environments of early SMGs \citep[e.g.,][]{baugh05, dwek11, hayward11,hayward13,  ferrara16}.  Probing the nature of these systems is, however, observationally expensive \citep[e.g.,][]{hodge12} and somewhat complex. Although  gravitationally lensed sources might overcome the issue of time-consuming follow-up observations \citep{strandet16, harrington16, harrington18} their derived quantities  might be affected by lens modeling uncertainties \citep{bussmann13}. \\

To investigate the role of this relatively unexplored SMG population in the context of galaxy formation and evolution,  \cite{smolcic15}  have presented  the largest sample of \emph{spectroscopically confirmed and unlensed} $z>4$ SMGs (six sources) in the  Cosmic Evolution Survey (COSMOS) field, where only AzTEC3 ($z=5.298$) --  studied in detail by  \cite{riechers10, riechers14} -- was excluded. To}  explore the dust distribution and gas kinematics,  high-resolution [CII]\,158\,$\mu$m line observations  have been secured towards three of these sources; namely, J1000$+$0234 ($z=4.544$), Vd$-$17871 ($z=4.624$) and AzTEC/C159 ($z=4.567$, Karim et al. in prep; hereafter K18). Based on those observations, \citet[hereafter J17]{jones17}   revealed  a gas dominated rotating disk in J1000$+$0234 and AzTEC/C159.  The  latter emerges as the best example of  a flat rotation curve at large radius.   These extreme systems, with gas dominated rotating disks and concomitant intense star formation activity, seem to  have no analog at lower redshifts; rendering their detailed study paramount to understand the formation of galaxies in the early Universe. \\

Here, we report the detection of $^{12}$CO(2$\to$1) and $^{12}$CO(5$\to$4)  line emission to  investigate the molecular gas properties of AzTEC/C159. $^{12}$CO observations have proved to be well suited to unveil the nature of the star-forming gas  of high-redshift galaxies \citep[e.g.,][]{frayer98, riechers08, schinnerer08, carilli10, engel10,riechers11, hodge12, hodge13, bothwell13, bouche13, carilli13,  debreuck14, sanchezalmeida14,  leroy15, narayanan15}. 
 Low-$J$ $^{12}$CO emission lines provide tight constraints on the mass and extent of the molecular gas reservoir, as well as the star formation efficiency (SFE).  In combination with {\rm  multiple}  high-$J$ $^{12}$CO line detections, it is possible to explore the $^{12}$CO spectral line energy distribution (SLED) and unveil the physical properties of the star-forming gas \citep[e.g., ][]{weiss05, weiss07,  carilli10, riechers10,  papadopoulos12, daddi15}.  {\rm While the $\rm ^{12}CO(2\to1)$ and $\rm ^{12}CO(5\to4)$ line  detections alone can not fully constrain the shape of the  $^{12}$CO SLED of AzTEC/C159,   they do  allow us  to estimate the brightness temperature ratio $r_{52}$ \cite[e.g.,][]{bothwell13}. This parameter  can be used to obtain initial  insights about the overall excitation conditions of the molecular gas in high-redshift  systems \citep[e.g.,][]{riechers10}.   Molecular line spectroscopic studies, on the other hand, can be well complemented with FIR photometric information tracing the dust obscured star-formation activity in SMGs; via FIR SED fitting  it is possible to estimate dust mass,  infrared luminosity ($L_{\rm IR}$) and hence star formation rates \cite[SFR, e.g. ][]{swinbank14}.} \\

After introducing AzTEC/C159 in Section \ref{aztecc159}, we present the details of the observations in Section \ref{observations}, followed by the results and discussion in Section \ref{results} and \ref{discussion}. A summary is given in Section \ref{summary}.  Throughout, we assume the following cosmology  $h_{0}= 0.7$, $\Omega_M=0.3$, and $\Omega_\Lambda=0.7$.

\begin{figure*}
	\begin{center}
		\includegraphics[width=12cm]{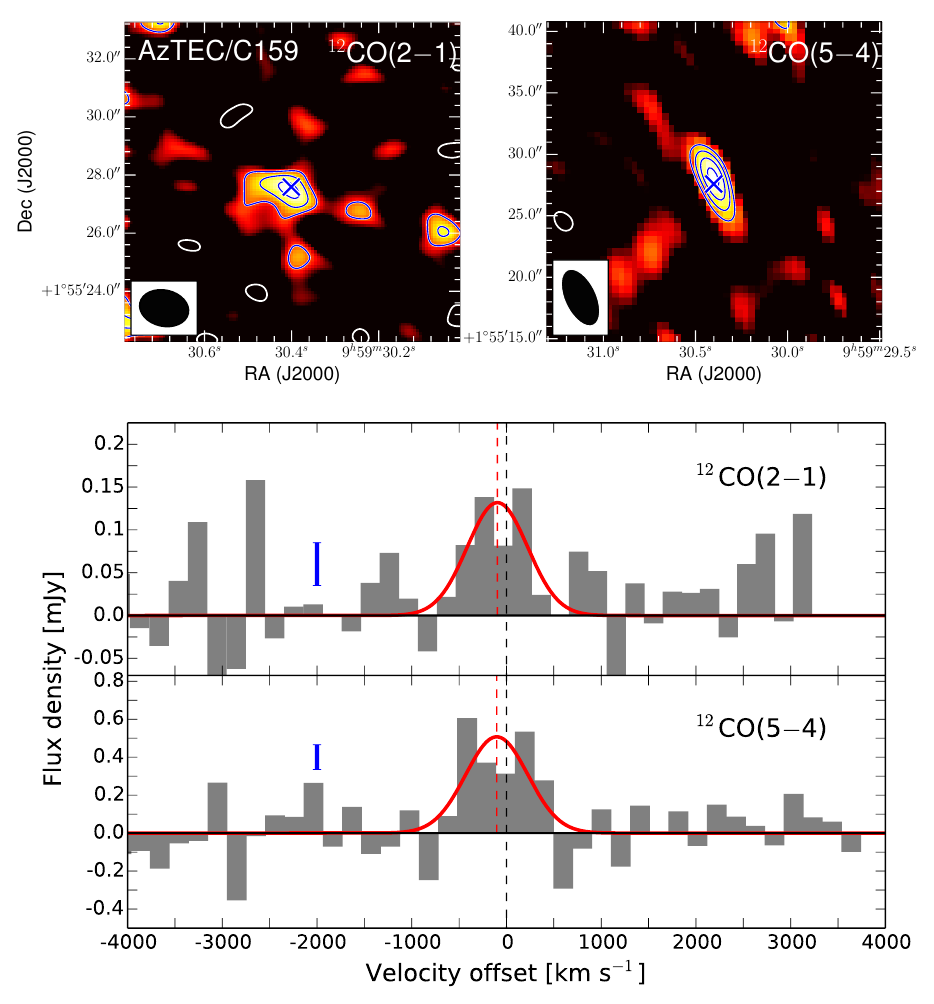}
		\caption{	{\it (Upper panels)}  
			The velocity-integrated intensity maps of the $^{12}$CO(2$\to$1) and $^{12}$CO(5$\to$4) line. We integrated  emission within the velocity range of $\pm750$\,km\,s$^{-1}$, i.e. $\sim8$ channels, where significant line emission was detected. The contour levels of the $^{12}$CO(2$\to$1) line emission map  are at [-2, 2, 3, 4]\,$\times\,1\sigma$, with $\sigma=0.05$\,mJy\,beam$^{-1}$ per 200\,km\,s$^{-1}$ wide channel. Negative contours are shown in white and positives in blue. The beam size of $1.70\times1.24$ arcsec is plotted on the bottom left corner. The blue cross denotes the peak position of the [CII]~158\,${\rm \mu m}$ emission line.  In the right panel we show the velocity-integrated intensity map of the $^{12}$CO(5$\to$4) line emission. Contour levels  are at [-3, 3, 4, 5, 6]\,$\times\,1\sigma$, with $\sigma=0.13$\,mJy\,beam$^{-1}$ per 200\,km\,s$^{-1}$ wide channel. The beam size of $5.0\times2.6$ arcsec is also plotted on the bottom left corner.
		{\it (Lower panels)}	The $^{12}$CO(2$\to$1) and $^{12}$CO(5$\to$4) line spectra binned to 200\,km\,s$^{-1}$ channels. The red-solid  lines are Gaussian fits to the line spectra, the central position  is shown by the red-dashed lines.   The blue error bar denotes the typical $1\sigma$\,rms value.  }	
		\label{colines}
	\end{center}
\end{figure*}

\section{AzTEC/C159}\label{aztecc159}
AzTEC/C159 was originally detected at the 3.7$\sigma$ level in the ASTE/AzTEC-COSMOS 1.1\,mm survey of the inner COSMOS 1\,deg$^2$ \citep{aretxaga11}.  Long-slit spectroscopy with DEIMOS/Keck has revealed a narrow Ly$\alpha$ line, which puts AzTEC/C159 at a redshift  $z=4.569$ \citep{smolcic15}.  The broadband spectral energy distribution (SED) from the optical through the FIR is consistent with a dusty star-forming galaxy, with a star formation rate (SFR) of $\sim700\,\text{M}_{\odot}\text{yr}^{-1}$,  stellar mass of $(4.5\pm0.4)\times10^{10}\,\text{M}_\odot$ and dust mass of $2.0^{+3.0}_{-1.2}\times10^{9}\,\text{M}_\odot$  \citep[][G\'omez-Guijarro et al.  submitted; hereafter GG18]{smolcic15}.  High-resolution [CII]~158\,$\mu$m line observations $\rm  (FWHM=0.36\,arcsec)$ with the Atacama Large Millimeter/submillimeter Array  (ALMA; project 2012.1.00978.S, PI: A. Karim) have revealed a double horn profile, while the [CII]~158${\rm \mu m }$ first moment and channel maps  {\rm are well described by} the classic pattern for rotating disks (J17, K18). AzTEC/C159 complements the sample of galaxies at $z>4$  exhibiting regular gas rotation on kpc-scales (i.e. disks); GN20 ($z=4.05$), ALESS73.1 ($z=4.755$), J1000$+$0234 ($z=4.542$), Vd$-$17871 ($z=4.622$),  and J1319$+$0950  \citep[$z = 6.127$, ][K18]{hodge12,debreuck14, jones17, shao17}. {\rm The  kinematic signatures of rotation revealed by ALMA (J17, K18), including  evidence of a flat rotation curve at large radius (J17),  renders AzTEC/C159  one of the best examples to date for an apparently rotating disk galaxy in the early Universe. We note, however, that given the achieved \hbox{$\sim2.3$\,kpc} resolution ALMA observations we can not completely rule out a merging scenario from the [C\,II] line dynamical analysis alone. In this context, studies that probe the physical properties of the ISM, e.g. molecular gas,  are  also key to indirectly probe the dominant mode of star-formation  in this system. }\\

\section{Karl G. Jansky VLA and NOEMA Observations}\label{observations}

$^{12}$CO(2$\to$1) line observations were carried out in January 2016
with the Karl G. Jansky Very Large Array (VLA) of the NRAO
in the D-- and DnC--configurations (project 15B-280, PI: A. Karim). Five observing
sessions,  3.5\,hrs each, resulted in a total of 17.5\,hrs --  with 9.0\,hrs on target. We used the quasar 3C286   for bandpass, delay, and flux calibration, and J1038+0512  for complex gain calibration. 
We made use of the Q-band receivers and the pair of 8-bit samplers on each VLA antenna, resulting in a pair of $1.024\,$GHz bands in right and left circular polarization. These bands were overlapped by $128\,$MHz to correct for the loss of signal at their edges, so the total bandwidth was 1.92$\,$GHz (from $40.963$ to $42.883\,$GHz); covering the redshifted $^{12}$CO(2$\to$1) line at 41.41\,GHz, according to the prior redshift estimation  of $z=4.567\pm0.002$  based on the [CII]~158${\rm \mu m }$ line detection with ALMA (K18).
Data were calibrated using the Common Astronomy Software Applications (CASA). Images were created with the CASA task \texttt{clean} and using a range of robust parameters. Ultimately, we use the image computed with robust=1 as it provides us with the best balance between spatial resolution and rms noise.  This results in a data cube with a synthesized beam size of  $1.70\times1.24$\,arcsec resolution  (PA\,$=77.5^\circ$)  with an rms of 0.05\,mJy\,beam$^{-1}$ for 27 MHz ($\sim$ 200\,km\,s$^{-1}$) wide channels.\\

The $^{12}$CO(5$\to$4) line was observed with the NOrthern Extended Millimeter Array (NOEMA) interferometer on December 8, 2015 and January 19, 2016  over two tracks (project 15CX001, PI: A. Karim) in D- and C-configuration.  We used the WideX correlator covering a frequency range of 3.6\,GHz. The tuning frequency of 103.51\,GHz was chosen to encompass the redshifted $^{12}$CO(5$\to$4) line, considering $z=4.567\pm0.002$.
We used the quasars 3C84 and B0906+015 as flux and phase/amplitude calibrators, respectively. The data calibration and mapping were performed with the  IRAM GILDAS software packages \texttt{clean} and \texttt{mapping}. The final cube corresponds to 9.4\,hrs on source, out of $\sim$19 hrs of total observing time; with a synthesized beam size  of $5.0\times2.6$ arcsec and PA$=23.5^\circ$ (using natural weighting).   We reach a sensitivity of 0.13\,mJy\,beam$^{-1}$ per 200\,km\,s$^{-1}$ wide channel.\\

\section{Analysis and Results}\label{results}
\subsection{CO lines}\label{coline}
The $^{12}$CO(2$\to$1) and $^{12}$CO(5$\to$4) integrated lines are detected at the 5.5 and 8.1$\sigma$ level, respectively. Fig. \ref{colines}  shows the intensity maps integrated over $\sim$1500\,km\,s$^{-1}$ where significant line emission was detected. The peak position of the $^{12}$CO lines spatially coincides with that of the [CII]~158\,${\rm \mu m}$ line (K18)  within the positional uncertainties of $\sim$0.2 and 0.5 arcsec, respectively.  {\rm The $^{12}$CO(2$\to$1) line  spectrum} was extracted within an ellipse of  $2.1\times1.7$\,arcsec (PA=90\,degs). This aperture contains the total extent of the [CII]~158\,${\rm \mu m}$ line emission ($\sim$7\,kpc; J17, K18) and  assures that no flux is missed. We fit Gaussians to the $^{12}$CO line spectra within $\pm$850\,km\,s$^{-1}$ of the centroid (see Fig.  \ref{colines}) and measure a {\sc FWHM} of 750\,km\,s$^{-1}$ for both lines. The line emission of $^{12}$CO(2$\to$1) is  centered at $41.425\pm0.02$\,GHz, giving $z=4.565\pm0.003$; while the $^{12}$CO(5$\to$4) centroid at $103.550\pm0.05$\,GHz yields $z=4.565\pm0.003$. These values are in agreement with the redshift derived from the [CII]~158\,${\rm \mu m}$ line of  $4.567\pm0.002$ (K18). Finally, we measure an integrated flux of $105\pm 19$\,mJy\,km\,s$^{-1}$ for the $^{12}$CO(2$\to$1) line and  $417\pm51$\,mJy\,km\,s$^{-1}$ for $^{12}$CO(5$\to$4).  We verify that the latter values differ by less than 6\% from those derived by adding individual flux densities per channel, within the velocity range used in the Gaussian fit. \\

The  measured $^{12}$CO integrated line flux of high-redshift galaxies is influenced by the CMB emission. 
While the higher temperature of the CMB at $z=4.5$ enhances  the line excitation, the background against which the $^{12}$CO lines are measured also increases \citep[e.g, ][]{dacuhna13}. 
For example, in a dense ISM ($n_{\rm H_2}=10^{4.2}{\rm cm^{-3}}$), with a gas kinetic temperature of $T_{\rm kin}=40$\,K,  we can measure  70\% and 80\% of the intrinsic value of the  $^{12}$CO(2$\to$1)  and $^{12}$CO(5$\to$4) line, respectively.
  In low density environments ($n_{\rm H_2}=10^{3.2}{\rm cm^{-3}}$) with $T_{\rm kin}=18$\,K   only  20\% of the intrinsic value of the  $^{12}$CO(2$\to$1) line is detected, and less than that for $^{12}$CO(5$\to$4)  \citep{ dacuhna13}.  Unfortunately,  we  lack estimates for the gas density and  kinetic temperature in AzTEC/C159.  Yet,  it is known that  high SFR is likely to be associated with  high temperatures of gas and dust as well as gas density \citep[e.g., ][]{narayanan14}. Under the assumption of thermodynamic equilibrium between the gas and dust, we would have $\hbox{$T_\mathrm{kin}$}\simeq\hbox{$T_\mathrm{dust}$}=37$\,K (from our mid-IR$-$mm SED fitting presented in Section \ref{firproperties}). Therefore, we favor the scenario with dense gas and $T_{\rm kin}=40$\,K that 
yields an increase in $S_{\text{CO(2$\to$1)}}$  and $S_{\text{CO(5$\to$4)}}$ by a factor [1/0.7] and [1/0.8], respectively \citep{dacuhna13}. Consequently, we derive  $S^{\rm intrinsic}_{\text{CO(2$\to$1)} }=150\pm 27$\,mJy\,km\,s$^{-1}$ and $S^{\rm intrinsic}_{\text{CO(5$\to$4)}}=521\pm 64$\,mJy\,km\,s$^{-1}$. \\

\begin{figure}[h]
	\begin{center}
		\includegraphics[width=9.cm]{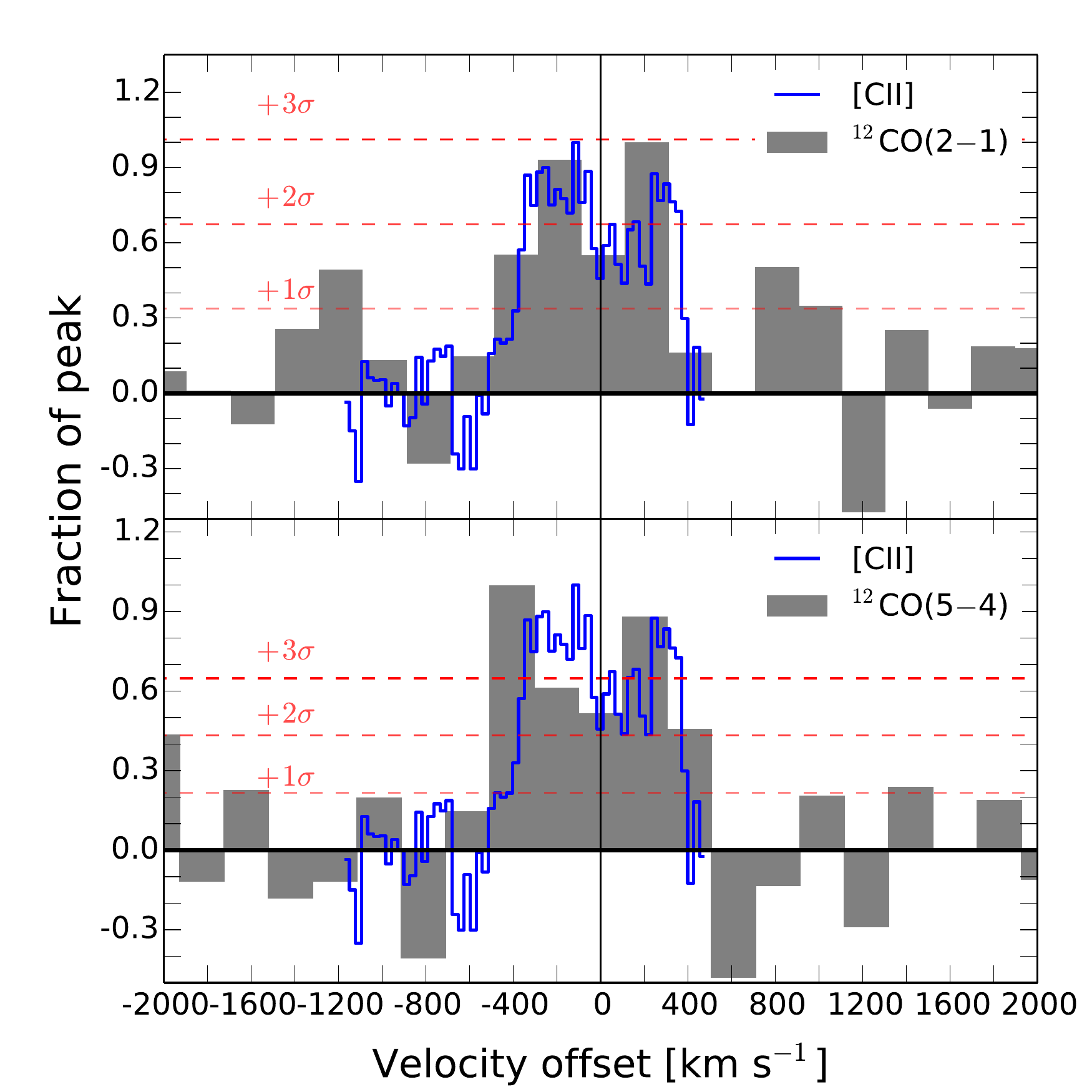}
		\caption{The [CII]~158\,${\rm \mu m}$ line spectrum of AzTEC/C159 (blue line; J17, K18) together with the $^{12}$CO (2$\to$1) and $^{12}$CO (5$\to$4)  detections from VLA and NOEMA, respectively.  Each line is renormalized by its peak intensity. Red dashed lines show the typical  noise level per channel. }	
		\label{profiles}
	\end{center}
\end{figure}

In Fig. \ref{profiles}, we compare the $^{12}$CO and [CII]~158\,${\rm \mu m}$ line profiles. We find that the $^{12}$CO lines are as broad as the [CII]~158\,${\rm \mu m}$ line (J17, K18), all with \hbox{{\sc FWHM}~$\sim750$~km\,s$^{-1}$}. There is also  marginal evidence of  double-horn profiles in both $^{12}$CO lines, where two peaks are   close/above the $3\sigma$ level.  {\rm To constrain the size of the $\rm ^{12}CO(2\to1)$ line emitting region  we fit a single Gaussian model to the uv data and find evidence for a compact/point-like structure.  
	Based on the synthesized beam size and SNR of the line detection, we use the results from \citet[][Eq. 7]{martividal12} to  estimate an upper limit of $\sim$1\,arcsec  for  the source size  --  corresponding to an intrinsic size of $\sim6.5$\,kpc at the redshift of AzTEC/C159.} The presence of an extended and possibly rotating molecular gas reservoir would be consistent with the  [CII]~158\,${\rm \mu m}$ gas extent ($\sim$7\,kpc) and dynamics of AzTEC/C159 (J17, K18).  Yet, we advise caution when interpreting our results on the $^{12}$CO size and dynamics of AzTEC/C159 due to  the current limitation of our $^{12}$CO line observations. \\

\subsection{FIR properties}\label{firproperties}
Continuum emission from AzTEC/C159 was detected at 3\,mm from our NOEMA observations. By averaging  line-free channel maps, around the $^{12}$CO(5$\to$4) centroid at 103.550\,GHz, we derive a value of $220\pm40$\,$\mu$\,Jy. No continuum emission was detected in our VLA observations at \hbox{$\sim7$\,mm}; based on the noise level of the continuum image,  we derive a $3\sigma$ upper limit of $\,20$\,$\mu$\,Jy. These new mm data points  mitigate the uncertainties in determining the FIR properties of AzTEC/C159. \\

\begin{figure}
	\begin{center}
		\includegraphics[width=8.7cm]{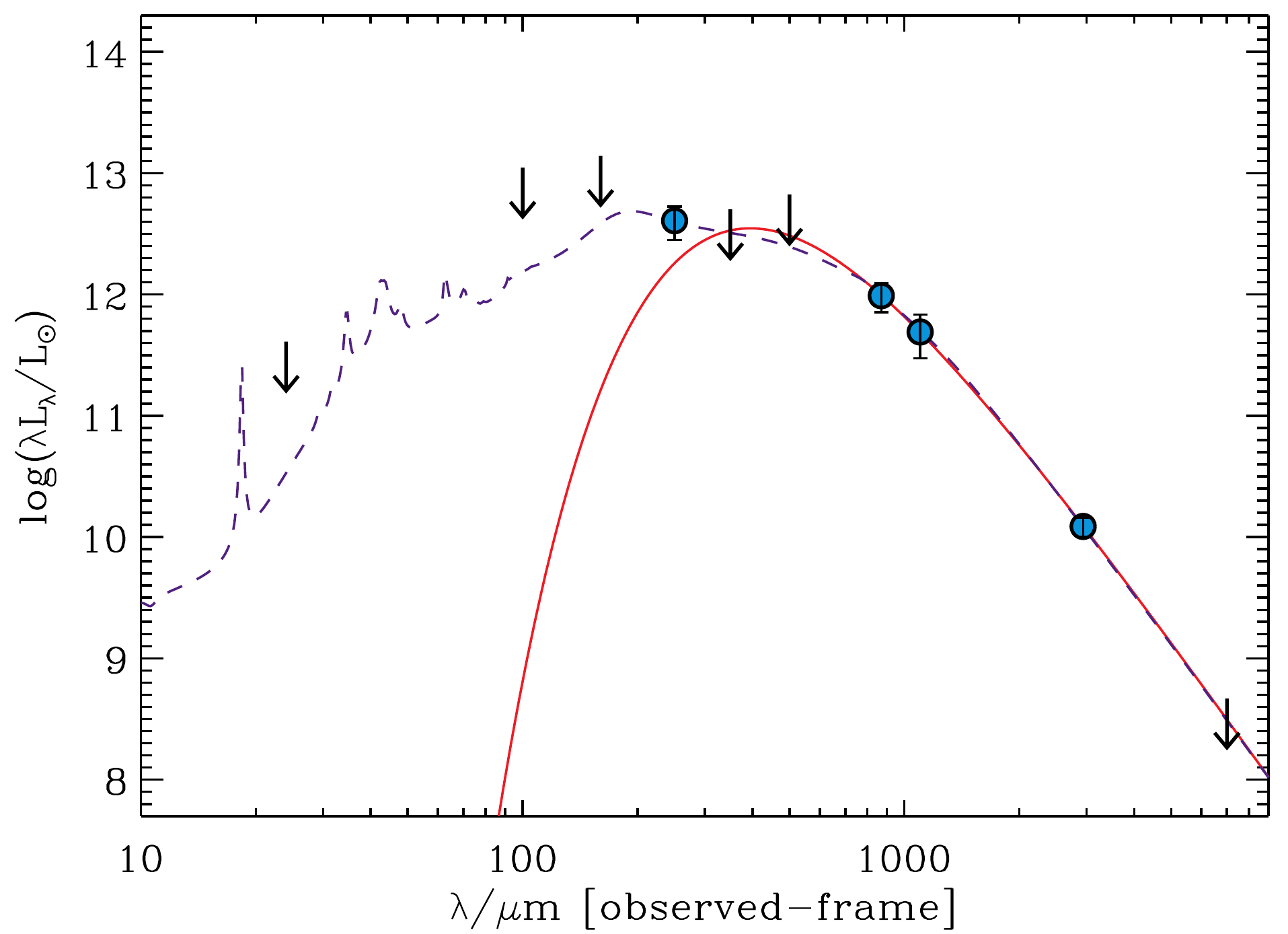}
		\caption{ Broadband SED of AzTEC/C159. The  best fit  \cite{draine07b} model  is shown by the blue dashed line.
			A modified blackbody  (red line) is used to fit the FIR-to-mm data points. 
			The monochromatic flux densities from {\it Herschel} PACS/SPIRE and JCMT/AzTEC observations used in the fit are listed as follows: $S_{100\mu{\rm m}}<6.8$\,mJy,   $S_{160\mu{\rm m}}<13.6$\,mJy, $S_{250\mu{\rm m}}=6.2\pm1.9$\,mJy, $S_{350\mu{\rm m}}<10.8$\,mJy, $S_{500\mu{\rm m}}<20.4$\,mJy and $S_{1.1{\rm mm}}=3.3\pm1.3$\,mJy \citep{smolcic15}. We add three more sub-mm/mm data points from our observations: $S_{870{\rm \mu m}}=6.9\pm0.2$\,mJy  (GG18), $S_{3{\rm mm}}=220\pm40$\,$\mu$\,Jy and $S_{7{\rm mm}}<20$\,$\mu$\,Jy.
			The downward pointing arrows denote upper limits to the corresponding flux densities. 	}	
		\label{sedfit}
	\end{center}
\end{figure}

We derive the infrared luminosity ($L_{\rm IR}$; in the range $8-1000\,\mu$m), SFR and  dust mass ($M_{\rm dust}$) of AzTEC/C159 via mid-IR$-$mm SED fitting {\rm using the \cite{draine07b} dust model}.  We refer the reader to \cite{smolcic15} for a detailed description of the fitting process.  As shown in Fig. \ref{sedfit},  aside from our mm SED constraints, the 1.1\,mm data point from the  JCMT/AzTEC 1.1\,mm  survey \citep{scott08} and ALMA 870\,$\mu$m continuum data point (GG18), we  use observations from the {\it Herschel} Space Observatory \citep{pilbratt10} towards the COSMOS field: those from the  Photodetector Array Camera and Spectrometer \citep[PACS at 100 and 160\,$\mu$m; ][]{lutz11} as well as the Spectral and Photometric Imaging Receiver \citep[SPIRE at 250, 350 and 500\,$\mu$m; ][]{oliver12}. 
We find $L_{\rm IR}=7.4^{+2.1}_{-1.7}\times10^{12}\,{\rm L}_\odot$, $T_{\rm dust}=37\pm3$\,K, $M_{\rm dust}=2.5^{+0.6}_{-0.5}\times10^{9}\,{\rm M}_\odot$; by assuming a Chabrier IMF and  the relation SFR [M$_{\odot}$yr$^{-1}$]=$10^{-10}L_{\rm IR} [{\rm L}_{\odot}]$   \citep{kennicutt98}, we estimate a SFR=$740^{+210}_{-170}\,{\rm M}_{\odot}{\rm yr}^{-1}$. {\rm The derived $L_{\rm IR}$ and  $T_{\rm dust}$ are 0.3\,dex and 5\,K larger, respectively, than the characteristic values for the full ALESS SMG sample with a median redshift of $\sim$2.5 \citep{swinbank14}. Similar to the $z>4$ SMGs presented by \cite{smolcic15},  AzTEC/C159 lies at the high-end of  the $L_{\rm IR} -T_{\rm dust}$  correlation \citep[e.g.,][]{chapman05, magnelli12}, consistent with an extrapolation of the trend for  {\it Herschel}-selected $0.1 < z < 2$ infrared galaxies \citep{symeonidis13}.   }\\

\begin{table*}
	\begin{center}
		\caption{Properties of AzTEC/C159}
		{\tiny
			\begin{tabular}{ l  l l l l }
				\hline \hline \\
				{\bf Properties} &  {\bf Units}  &  \multicolumn{2}{|c|}{\bf Values}  \\[0.5ex]
				&   & $^{12}$CO(2$\to$1)   &  $^{12}$CO(5$\to$4)  \\[0.5ex]
				\hline	\\
				FWHM & km\,s$^{-1}$ & 750 $\pm$ 200   & 770 $\pm$ 360\\[0.5ex]
				Peak flux  & mJy & 0.13 $\pm$ 0.05    & 0.50 $\pm$ 0.13\\[0.5ex]
				Integrated flux & mJy\,km\,s$^{-1}$ & 105 $\pm$ 19\tablefootmark{a}   &   417 $\pm$ 51\tablefootmark{a}  \\[0.5ex]
				Peak frequencies &   GHz  &   41.425 $\pm$ 0.020   & 103.550 $\pm$ 0.050 \\[0.5ex]
				\hline \\
				$z$\tablefootmark{b} &  $\dots$ &  \multicolumn{2}{c}{ 4.567$\pm$0.002}\\[0.5ex]
				RA, DEC\tablefootmark{b} & $\dots$ &  \multicolumn{2}{c}{ 09:59:30.401 +01:55:27.59} \\ [0.5ex]
				$L_{\text{IR}}$   & L$_{\odot}$  &      \multicolumn{2}{c}{7.4$^{+2.1}_{-1.7}\times 10^{12}$} \\[0.5ex]
				SFR   & M$_{\odot}$ yr$^{-1}$  &      \multicolumn{2}{c}{740$^{+210}_{-170}$} \\[0.5ex]
				$M_{\star}$\tablefootmark{c}   & M$_{\odot}$  &      \multicolumn{2}{c}{$(4.5\pm0.4)\times10^{10}$} \\[0.5ex]
				$M_{\text{dust}}$   & M$_{\odot}$  &      \multicolumn{2}{c}{$2.5^{+0.6}_{-0.5}\times10^{9}$} \\[0.5ex]
				$M_{\text{dyn}}$\tablefootmark{d}   & M$_{\odot}$  &      \multicolumn{2}{c}{$2.8^{+1.1}_{-0.6}\times10^{11}$} \\[0.5ex]
				\hline \\
				$S_{103\text{GHz}}$ & $\mu$\,Jy  &   \multicolumn{2}{c}{220$\pm$40} \\[0.5ex]
				$S_{41\text{GHz}}$& $\mu$\,Jy  &   \multicolumn{2}{c}{$<$\,20} \\[0.5ex]
				$L'_{\text{CO}}$\tablefootmark{e} &  K\,km\,s$^{-1}$\,pc$^{2}$ &   \multicolumn{2}{c}{$(3.4\pm0.6)\times10^{10}$}\\[0.5ex]
				$M_{\rm H_2}$($\alpha_{\text{CO}}/4.3)$\tablefootmark{e}  &  M$_\odot$ &  \multicolumn{2}{c}{$(1.5\pm0.3)\times10^{11}$}   \\[0.5ex]
				$\tau_{\text{gas}}\times(\alpha_{\text{CO}}/4.3)$\tablefootmark{e}         &   Myr &  \multicolumn{2}{c}{$200\pm100$} \\[0.5ex]
				$\mu_{\text{gas}}\times(\alpha_{\text{CO}}/4.3)$\tablefootmark{e}         &  $\dots$ &  \multicolumn{2}{c}{$3.3\pm0.7$} \\[0.5ex]
				$L_{\text{IR}}/L'_{\text{CO}}$ & 
				 L$_{\odot}$(K\,km\,s$^{-1}$\,pc$^{2}$)$^{-1}$  & \multicolumn{2}{c}{ $216\pm80$}  \\[0.9ex]
				\hline 
				\hline
			\end{tabular}
		     \tablefoot{
			\tablefoottext{a}{After considering the effect of the CMB,  the integrated flux density of $^{12}$CO(2$\to$1)   and $^{12}$CO(5$\to$4)   will increase by a factor [1/0.7] and [1/0.8], respectively (see Section \ref{coline}). We use the corrected value, i.e. $S^{\rm intrinsic}_{\text{CO(2$\to$1)}}=150\pm 27$\,mJy\,km\,s$^{-1}$ and $S^{\rm intrinsic}_{\text{CO(5$\to$4)}}=521\pm 64$\,mJy\,km\,s$^{-1}$.}
			\tablefoottext{b}{Karim et al. (in prep)}
			\tablefoottext{c}{G\'omez-Guijarro et al. (submitted)}		\tablefoottext{d}{\cite{jones17}; within a radius of 2.9\,kpc}
			\tablefoottext{e}{Values corrected from the CMB.}}}
	\end{center} 
\end{table*}

\subsection{ Molecular gas mass and SFE}

The molecular gas mass can be estimated from the relation:
 \begin{equation}\label{eq:1}
  M_{\rm H_2}=\alpha_{\text{CO}} L'_{\text{CO}}~,
 \end{equation}
\noindent
 where $\alpha_{\text{CO}}$ is the CO luminosity to H$_2$ mass conversion factor and $L'_{\text{CO}}$ is the $^{12}$CO(1$\to$0) line luminosity. The latter can be derived from our $^{12}$CO(2$\to$1) line assuming that  $L'_{{\rm CO}}=1.18 L'_{{\rm CO}(2\to1)}$  for typical SMG excitation conditions (consistent with the results in Section \ref{excitation})  and using the relation  \citep{carilli13}:

 \begin{equation}
 L'_{{\rm CO}(2\to1)}=3.25\times 10^7 S_{\text{CO(2$\to$1)}} \nu^{-2}_{\text{CO(2$\to$1)}}D^2_{\text{L}}(1+z)^{-1},
 \label{eq:cw}
 \end{equation}
 
\noindent
 where $S_{\text{CO(2$\to$1)}}$ is the integrated line flux in Jy\,km\,s$^{-1}$, $\nu_{{\rm CO}(2\to1)}$ is the rest-frame frequency of the $^{12}$CO(2$\to$1)  line in GHz and $D_{\text{L}}$ the luminosity distance in Mpc.  Based on $S^{\rm intrinsic}_{\text{CO(2$\to$1)}}$ (i.e. the corrected value from the CMB) and Eq. \ref{eq:cw}, we derive 	$L'_{\text{CO}}= (3.4\pm0.6)\times10^{10}$\,K\,km\,s$^{-1}$\,pc$^{2}$.

 The value  for $\alpha_{\text{CO}}$ depends on local interstellar medium (ISM) conditions and, consequently, may vary across different galaxy types  \citep{daddi10, papadopoulos12, magnelli12,bolatto13}. {\rm  According to \cite{papadopoulos12},  the major driver of the $\alpha_{\rm CO}$ value is the average dynamical state of the molecular gas. As a result, while } low values for $\alpha_{\rm CO}$ could be related to highly turbulent molecular gas -- likely associated  with merging systems -- as observed in local ultra-luminous infrared galaxies (ULIRGs) with $\alpha_{\text{CO}}=0.8$\,M$_{\odot}$\,K$^{-1}$\,km$^{-1}$\,s\,pc$^{-2}$  \citep[e.g., ][]{downes98},  self-gravitating gas yields high $\alpha_{\rm CO}$ values \citep[e.g., ][]{papadopoulos12b} as in star-forming spiral disks like the Milky-Way with $\alpha_{\text{CO}}=4.3$\,M$_{\odot}$\,K$^{-1}$\,km$^{-1}$\,s\,pc$^{-2}$  \citep[e.g.,][]{abdo10}. {\rm  Spatially resolved observations of nearby SFGs have also probed  that  $\alpha_{\rm CO}$ might vary within the galaxy itself, where central values could be significantly smaller than the galaxy average  \citep{sandstrom13}.  For high-redshift unresolved sources, on the other hand,  we have to rely on an average $\alpha_{\rm CO}$ that reflects the overall physical conditions of the molecular gas. Since J17 and K18 have revealed a rotationally supported gas disk in AzTEC/C159, it is likely that the bulk of the molecular gas is in self-gravitating clouds, pointing towards a Milky Way-like $\alpha_{\rm CO}$ value. We discuss in more detail the nature of $\alpha_{\rm CO}$ in this source and validate this assumption in Section 4.5.  By adopting $\alpha_{\text{CO}}=4.3$\,M$_{\odot}$\,K$^{-1}$\,km$^{-1}$\,s\,pc$^{-2}$ we find}  $M_{\rm H_2}=(1.5\pm0.3)\times10^{11}\,{\rm M}_\odot$. Together with the stellar mass of AzTEC/C159 of $M_{\star}=(4.5\pm0.4)\times10^{10} {\rm M}_\odot$ (GG18), we estimate a gas fraction  of $\mu_{\rm gas}\equiv M_{\rm H_2}/M_{\star}=3.3\pm0.7$ that is five times larger than the median for Main Sequence (MS) galaxies at $z\sim3$ \citep{schinnerer16}.   It should be noted that  even when assuming a ULIRG-like $\alpha_{\rm CO}$ prescription (and a CMB uncorrected $S_{\rm CO(2\to1)}$ flux),  the gas fraction would be high, i.e. $0.43\pm0.10$. The massive molecular gas reservoir of AzTEC/C159 is consistent with the general picture of having large cool-ISM reservoirs  fueling intense star formation activity at high redshift \citep[e.g.,][]{carilli13}. \\

\begin{figure*}[h!]
	\begin{center}
		\includegraphics[width=14.cm]{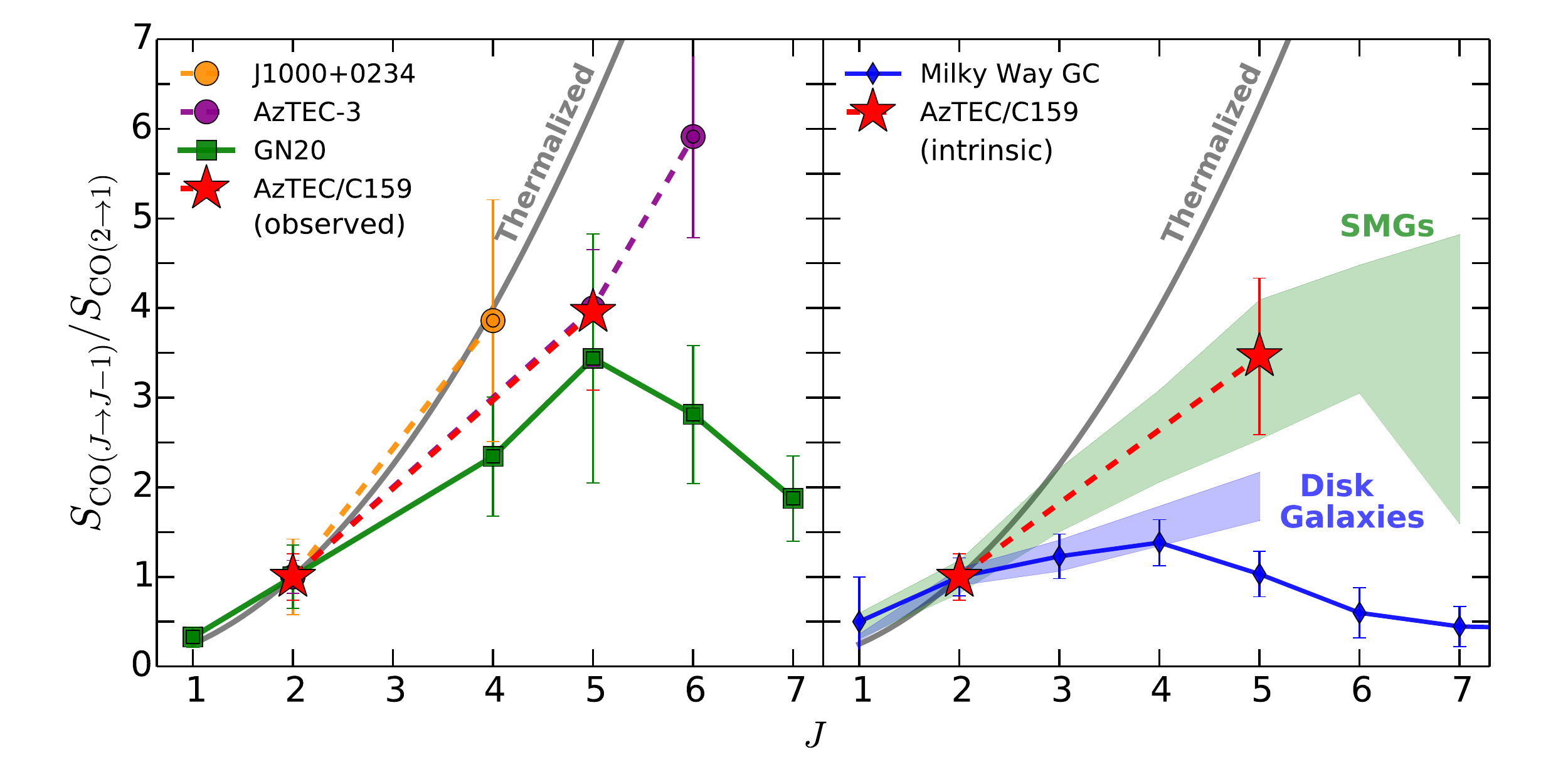}
		\caption{ ({\it Left panel})  Observed $^{12}$CO SLED of AzTEC/C159 and others $z>4$ SMGs for comparison: GN20 \citep{carilli10}, AzTEC-3 \citep{riechers10} and J1000$+$0234 \citep{schinnerer08}. Grey lines show the expected trend for thermalization, i.e. LTE conditions. Dashed/solid color lines connect  the data points  to illustrate the shape of the different $^{12}$CO SLEDs. The line fluxes have been normalized to the $^{12}$CO(2$\to$1) line.  Given the high-redshift nature of these systems, it is likely that the CMB emission affects the shape of the  $^{12}$CO SLEDs \citep{dacuhna13}. ({\it Right panel}) Intrinsic  $^{12}$CO SLED of AzTEC/C159. The observed flux densities for $^{12}$CO(2$\to$1) and $^{12}$CO(5$\to$4) have been corrected up by 1/0.7 and 1/0.8, respectively, to take into account the effect of the CMB  (see Section \ref{coline}).  We also present the $^{12}$CO SLED of the 	Milky Way galactic center (GC) \citep{fixsen99}, MS disk-galaxies at $z\sim1.5$ \citep{daddi15}  and a sample of SMGs \citep[median of $z=2.2$, ][]{bothwell13};  {\rm  similarly, we normalize all the line  fluxes  to the $\rm ^{12}CO(2\to1)$ line.} Shaded regions show the dispersion of the $^{12}$CO SLEDs for different SMGs and disk galaxies samples. At the redshift of the latter sources, the impact of the CMB on the observed $^{12}$CO line flux densities is negligible \citep{dacuhna13}. Both the observed and intrinsic  $S_{\rm CO(5\to4)}/S_{\rm CO(2\to1)}$ ratios of AzTEC/C159, are  consistent with high gas excitation condition.   }		
		\label{cosed}
	\end{center}
\end{figure*}

We use the {\rm empirical ratio} $L_{\rm IR}/L'_{\rm CO}$ as a proxy of the SFE of AzTEC/C159, {\rm allowing a more direct comparison with the average SFE of different galaxy populations.}  Based on the information in Table 1, we derive $L_{\text{IR}}/L'_{\text{CO}}=(216\pm80)\,{\rm L}_{\odot}$(K\,km\,s$^{-1}$\,pc$^{2}$)$^{-1}$, placing  it within the upper scatter of the $L_{\text{IR}}/L'_{\text{CO}}$ correlation as presented by \cite{carilli13}. This ratio is much higher than that observed in normal star-forming disk galaxies such as nearby spirals and $z\sim(0.4-2)$ MS galaxies \citep[between 20 and 100 L$_{\odot}$(K\,km\,s$^{-1}$\,pc$^{2}$)$^{-1}$; ][]{daddi10,  magnelli12} but comparable to merger-driven starbursts such as  local ULIRGs and SMGs   \citep[e.g.,][and references therein]{genzel10, bothwell13, aravena16}. 
The SFE of  AzTEC/C159 is, however,  similar to those of the very high-redshift starburst disk galaxies GN20 \citep[$\sim$180\,L$_{\odot}$(K\,km\,s$^{-1}$\,pc$^{2}$)$^{-1}$; ][]{carilli10, hodge12}, and J1000$+$0234 \citep[$\sim$140\,L$_{\odot}$(K\,km\,s$^{-1}$\,pc$^{2}$)$^{-1}$; ][]{schinnerer08,smolcic15}. 
Thus, while the extreme SFE of AzTEC/C159 seems  unusual for a star-forming disk galaxy at $z<2$, at $z>4$  disk galaxies harboring vigorous SFE (like AzTEC/C159, GN20 and J1000$+$0234) might be common.

\subsection{Gas excitation}\label{excitation}

We combine our  $^{12}$CO(2$\to$1) and $^{12}$CO(5$\to$4) line detections   to obtain insights into the molecular gas excitation  in AzTEC/C159.  As mentioned before, at $z\sim4.5$ the CMB emission affects the observed  $^{12}$CO line flux density in both lines \citep{dacuhna13}. This could modify the  $^{12}$CO(5$\to$4)/CO(2$\to$1)  line brightness temperature ratio ($r_{52}$) and, in consequence, our interpretation. Let us first consider the observed line flux densities in AzTEC/C159 and compare its $^{12}$CO SLED with those of some SMGs at $z>4$, where  the effect of the CMB  might be considerable but no corrections have been applied yet.  In Fig. \ref{cosed}, we plot the  $^{12}$CO SLED of J1000$+$0234 \citep{schinnerer08} and GN20 \citep{carilli10}, both star-forming disk galaxies with extreme SFE.  We also show the  $^{12}$CO SLED of AzTEC-3, a massive starburst galaxy possibly triggered by a major merger and associated with a protocluster at $z=5.3$ \citep{riechers10}. All these $^{12}$CO SLEDs are consistent with high gas excitation, where $r_{52}=0.63\pm0.16$   for AzTEC/C159. 

The impact of the CMB on the $^{12}$CO SLED shape below $z=2.5$ is negligible \citep{dacuhna13}.  In consequence, no corrections need to be made to the observed median $^{12}$CO SLED of SMGs (median redshift of $z=2.2$ as in \cite{bothwell13}) and disk galaxies at $z\sim1.5$ \citep{daddi10}.  We apply corrections to the observed flux densities of AzTEC/C159  to mitigate potential bias due to the CMB emission (see Section \ref{coline}), and compare its ``intrinsic'' $^{12}$CO SLED with those of the aforementioned galaxy populations. As observed in Fig.~\ref{cosed}, with $r_{52}=0.55\pm0.15$, the molecular gas excitation conditions of AzTEC/C159 are consistent with those observed in SMGs, and significantly  deviates from  the expected trend for star-forming disk-like MS galaxies at  $z\sim1.5$ \citep{daddi15} and the MW  \citep{fixsen99}. The  elevated  $^{12}$CO SLED  of AzTEC/C159 might be a result of intrinsic processes, such as  elevated gas density and kinetic temperature that  yield a higher  rate of collisions between $^{12}$CO and H$_2$ \citep[e.g.,][]{narayanan14}. \\

 \subsection{The CO$\to$\,H$_2$ conversion factor  }

By determining the $\alpha_{\rm CO}$ value in AzTEC/C159, we can mitigate uncertainties on our gas mass estimates and, at the same time, infer the overall physical conditions of its molecular gas \citep[e.g.,][]{papadopoulos12b,papadopoulos12}. Here, we constrain  $\alpha_{\text{CO}}$  from  (a)  the gas-to-dust ratio ($\delta_{\text{GDR}}$) and (b)
the balance between the baryonic and dark matter content with the dynamical mass.

\subsubsection{The gas-to-dust method}
 By assuming that the molecular component dominates  the ISM, i.e.   $M_{\rm H_2}\gg M_\text{HI} + M_\text{HII}$, the molecular gas mass can  be estimated from the relation \citep[e.g., ][]{magnelli12, aravena16}:

\begin{equation}\label{eq:2}
	M_\text{gas}\simeq M_{\rm H_2}=\delta_{\text{GDR}}M_{\text{dust}}, 
\end{equation}

\noindent
where $\delta_{\text{GDR}}$ is the gas-to-dust ratio relation constrained locally by \cite{leroy11} that varies with metallicity as  $\log(\delta_{\text{GDR}})=9.4-0.85\times[12+\log(\text{O/H})]$. 
In the absence of direct metallicity estimates for AzTEC/C159, as well as a robust mass-metallicity relation at such high redshift, we assume that AzTEC/C159 has  solar metallicity. Then, by combining Eq. \ref{eq:1} and \ref{eq:2}:

\begin{equation}\label{eq:4}
	\alpha_{\text{CO}}=\frac{\delta_{\text{GDR}}M_{\text{dust}}}{L'_{\text{CO}}}. 
\end{equation}

\noindent
In Section \ref{coline}, we favored a scenario with dense gas and $T_{\rm kin}=40$\,K  to correct the observed $L'_{\text{CO}}$ by the effect of the CMB. Here, we also 
explore the possibility of having cool ($T_{\rm kin}=14$\,K) and low-density molecular gas, i.e. according to  \cite{dacuhna13} we vary $L'_{\text{CO}}$ within the range $[3.4\pm0.6,12.0\pm2.2]\times10^{10}$\,K\,km\,s$^{-1}$\,pc$^{2}$.
Based on \hbox{$M_{\text{dust}}=2.5^{+0.6}_{-0.5}\times10^9{\rm M}_{\odot}$} (derived in Section \ref{firproperties}) and Eq. \ref{eq:4} we estimate  \hbox{$\alpha_{\text{CO}}=3.8^{+2.6}_{-1.4}$\,M$_{\odot}$\,K$^{-1}$\,km$^{-1}$\,s\,pc$^{-2}$}. 
Note that a lower value of $\delta_{\text{GDR}}$ could lower the estimated $\alpha_{\text{CO}}$, but it would require  unlikely supra-solar metallicity.  On the contrary, sub-solar metallicities would yield even higher $\alpha_{\text{CO}}$.     \\

\subsubsection{The dynamical mass method}
Another approach to constrain  $\alpha_{\text{CO}}$ is based on the estimation of the mass content, which should match the baryonic and dark matter content, i.e.  $M_{\text{dyn}}=M_{\text{ISM}}+M_\star+M_{\text{DM}}$.  Since we assume that the ISM is dominated by molecular gas, then $M_{\text{ISM}}\simeq M_{\text{H2}}$. This,  in combination with Eq. \ref{eq:1}, leads to:

\begin{equation}\label{eq:5}
\alpha_{\text{CO}}=\frac{M_{\text{dyn}}-M_\star-M_{\text{DM}}}{L'_{\text{CO}}}.
\end{equation}

\noindent
The [CII]~158\,${\rm \mu m}$ gas dynamics of AzTEC/C159 has been characterized in  detail by J17.  Via the tilted ring fitting  method they find that at an intrinsic radius of 2.9\,kpc $M_{\text{dyn}}=(2.8^{+1.1}_{-0.6})\times10^{11}{\rm M}_\odot$. This value agrees within the uncertainties with the value derived by K18 ($M_{\text{dyn}}\sim2.4\times10^{11}{\rm M}_\odot$) using the software ``3D-Barolo''  and assuming an inclination angle of $\sim30$\,degs as in J17. Within an aperture radius of 2.9\,kpc, we find that $S^{\rm observed}_{\rm CO(2\to1)}=91\pm16$\,mJy, which is $\sim13$\% lower than that derived from  the aperture that contains the full emission line (see Section \ref{coline}). Consequently, as in the gas-to-dust ratio method, we consider $L'_{\text{CO}}\in[2.9\pm0.5,10.5\pm1.9]\times10^{10}$\,K\,km\,s$^{-1}$\,pc$^{2}$. \\

{\rm  GG18 have estimated the stellar mass of AzTEC/C159 ($[4.5\pm0.4]\times10^{10} {\rm M}_\odot$) using LePHARE \citep{arnouts99, ilbert06} and adopting the \cite{bruzual03}  stellar population synthesis models, a \cite{chabrier03}  IMF, and a exponentially declining star formation history (SFH). For this purpose, a substantial  optical/near-IR dataset was used by GG18: the  {\it g, r, i, z} and {\it y}
	bands from the Subaru Hyper Suprime-Cam imaging \citep[HSC, ][]{tanaka17}, near-IR J, H and Ks
	bands from  the UltraVISTA DR3 survey \citep{mccracken12}, and the mid-IR Spitzer/IRAC 3.6 and 4.5$\mu$m bands from the Spitzer Large Area Survey with
	Hyper-Suprime-Cam (SPLASH, Capak et al. in prep.).
	As discussed in detail in GG18, this stellar mass estimate could be subject to a number of  systematic uncertainties. 
	For instance, if the stars formed in-situ, it is possible that a  fraction of the  stellar light is being obscured by the dust and, therefore, the stellar mass used here might be underestimated	\footnote{\rm High spatial resolution rest-frame UV/optical/FIR observations are required to map the dust and stellar light distribution and better constrain the stellar mass of AzTEC/C159. This would also allow one to explore the presence of multiple interacting components (and not a massive single component) that might fit into a merger-driven star formation scenario. }. GG18 have quantified the stellar mass fraction we might be missing by this effect using different prescriptions for  the dust-to-stellar-mass ratio (DTS). By considering the ratio derived  from simulations in \citep{popping17} of $\log(\rm DTS) = -1.8$, the stellar mass would increase by  $\sim 0.4$\,dex; on the other hand,  the median DTS ratio for local ULIRGs $\log(\rm DTS) = -2.83$ \citep{calura17}  yields an increment of $\sim 1.1$\,dex.  The original estimate for the stellar mass (combined with the SFR of $740^{+210}_{-170}\,\rm M_\odot yr^{-1}$)  put AzTEC/C159  on the  upper-end of the MS of SFGs  at $z\sim4.5$ -- as given by \citep{schreiber15}. 
	An increase in stellar mass by $\sim0.4$\,dex would instead place AzTEC/C159 right on the MS of SFGs, while an increase by $\sim1.1$\,dex would lead to a very unlikely scenario in which AzTEC/C159 lies significantly below the MS. From this, we conclude that the stellar mass of AzTEC/C159 reported by GG18 might be underestimated by at most $\sim$0.4\,dex.
	By applying  Eq. \ref{eq:5} and considering the original estimate for the stellar mass ([$4.5\pm0.4]\times10^{10} {\rm M}_\odot$) from GG18, as well as a negligible dark matter component, we derive\footnote{A highly conservative error bar for $M_{\rm dyn}$ of 1\,dex  has been suggested by J17 to take into account systematic uncertainties of the method; i.e. $M_{\text{dyn}}=(2.8^{+6.1}_{-1.9})\times10^{11}{\rm M}_\odot$. This large range for $M_{\text{dyn}}$ also affords a wider range of inclination angles.  We derive $\alpha_{\text{CO}}=6.7^{+6.3}_{-4.4}$\,M$_{\odot}$\,K$^{-1}$\,km$^{-1}$\,s\,pc$^{-2}$. Thus,  what remains true is that $\alpha_{\text{CO}}$ is still higher than the prescription for ULIRGs-like systems  ($\alpha_{\text{CO}}=0.8$\,M$_{\odot}$\,K$^{-1}$\,km$^{-1}$\,s\,pc$^{-2}$).} $\alpha_{\text{CO}}=3.9^{+2.7}_{-1.3}$\,M$_{\odot}$\,K$^{-1}$\,km$^{-1}$\,s\,pc$^{-2}$.  The addition of a 0.4\,dex factor in the stellar mass  (i.e. $[1.0\pm0.4]\times10^{11} {\rm M}_\odot$) still yields a relatively high $\alpha_{\text{CO}}$ value of  $2.7^{+2.1}_{-1.1}$\,M$_{\odot}$\,K$^{-1}$\,km$^{-1}$\,s\,pc$^{-2}$. } \\

Our two independent methods point towards a high value of $\alpha_{\rm CO}$. This result agrees well with our previous assumption of $\alpha_{\text{CO}}\sim4.3$\,M$_{\odot}$\,K$^{-1}$\,km$^{-1}$\,s\,pc$^{-2}$.  
It is known that the most important influencing factor on $\alpha_{\text{CO}}$ is the average dynamical state of the molecular gas. While  low values are associated with unbound  gas, as observed in local ULIRGs, high values are related with self-gravitating gas \citep[e.g.,][]{papadopoulos12}; hence, the molecular gas dynamics in AzTEC/C159 might be more similar to that of local star-forming disks than those of disturbed major mergers.

\section{Discussion}\label{discussion}

Molecular line observations towards AzTEC/C159 reveal  a gas-rich system, elevated SFE, and high gas excitation. Its high $\alpha_{\rm CO}$ value  is consistent with a self-gravitating molecular gas distribution \citep[e.g.,][]{papadopoulos12b}.  In addition, [CII]~158\,${\rm \mu m}$ line observations have already revealed a gas dominated rotating disk extending up to a radius of $\sim$3\,kpc (J17). These dynamical properties do not fit into the scenario of gas-rich  galaxy mergers at high redshift  driving tidal torques that form dense gas regions (a  condition for high gas excitation) and trigger intense star formation activity \citep[e.g.,][]{hopkins06, hayward11, hayward12}. Then, 
the question arises as what would be  the physical mechanisms responsible for fueling and triggering such extreme star formation environment in this gas-rich rotating disk galaxy at $z=4.5$. \\

According to numerical simulations, massive rotating disk galaxies in the early Universe are not a rare phenomenon \citep[e.g.,][]{dekel09b, romano-diaz11}. In overdense environments, the smooth infall and accretion of cold gas from cosmological filaments can build up a disk \citep{feng15}.  This relatively smooth accretion, which dominates the mass input \citep{romano-diaz14, anglesalcazar17},  can  maintain an unstable dense gas-rich disk  that breaks into giant clumps forming stars at a high rate \citep{dekel09b}. Such star formation activity might be enhanced due to gravitational harassment, as  advocated for the star-forming disk galaxy at $z=4.05$ GN20 \citep{carilli10,hodge12}. \\

The latter scenario agrees with the high gas fraction and SFE of AzTEC/C159, but it is at odds with the high gas excitation condition  usually associated with major-mergers \citep[e.g.,][]{bournaud15}. However, as predicted by \cite{papadopoulos12b}, high-redshift disk-like systems with extreme SFE -- like AzTEC/C159 -- could also have high gas excitation conditions as a result of heating by turbulence and/or cosmic rays.  In fact,  by  $z\sim 3$  the mean radiation field intensity  $\langle$U$\rangle$ in MS galaxies might become similar to that of local ULIRGs and their $^{12}$CO SLEDs may look similar at this redshift \citep{daddi15}. We note that an elevated $^{12}$CO SLED could be also a result of AGN-driven mechanical and radiative feedback  \citep{papadopoulos08, papadopoulos10, dasyra14, moser16}; however, there is no robust evidence of an AGN in AzTEC/C159 \citep{smolcic15}. \\

It is  difficult to draw a coherent conclusion on galaxy formation and evolution from  this source alone. Nevertheless, AzTEC/C159 emerges as another gas-rich  disk galaxy at \hbox{$z\sim4$} with high SFE and  extreme $^{12}$CO SLED: together with J1000$+$0234 \citep[][J17]{schinnerer08} and GN20 \citep{carilli10, hodge12}. 
This  galaxy population  complements our picture on galaxy evolution. In particular,  \cite{toft17} suggest that such $z>4$ SMGs  with cool gas reservoirs distributed in rotating disks, and high SFE, might be the progenitors of massive quenched galaxies at $z\sim2$  that --surprisingly-- also exhibit a rotating disk \citep{vanderwel11, newman15}. To confirm this, we need to exploit  the synergy  between ALMA and the upcoming JWST, that promise a better understanding of the multi-phase gas and  stellar properties of early systems.

\section{Summary}\label{summary}
We reported  the molecular gas properties of the star-forming disk galaxy AzTEC/C159  at $z=4.567$. 
By using $^{12}$CO line observations of the transition levels   \emph{J}=2$\to$1 and \emph{J}=5$\to$4, and extensive ancillary data from the COSMOS collaboration, we have found that:

\begin{enumerate}
	\item The molecular gas mass of AzTEC/C159 is $M_{\rm H2}(\alpha_{\rm CO}/4.3)=(1.5\pm0.3)\times10^{11}\,\text{M}_\odot$, which  yields a high gas fraction of $\mu_{\rm gas}(\alpha_{\rm CO}/4.3)\equiv M_{\rm H_2}/M_{\star}=3.3\pm0.7$. 
	Its $L_{\text{IR}}/L'_{\text{CO}}$ ratio of $(216\pm80$) L$_{\odot}$\,(K\,km\,s$^{-1}$\,pc$^{2}$)$^{-1}$, i.e. SFE,  is comparable with that of local ULIRGs and SMGs \citep[e.g., ][and references therein]{aravena16}. \\
	
	\item The $^{12}$CO lines show tentative evidence of a double-peak line profile. Their {\sc FWHM} is comparable to that of the [CII]~158\,${\rm \mu m}$ line ({\sc FWHM}\,$\sim750$\,km\,s$^{-1}$; J17, K18), that is consistent with an extended gas reservoir.  However, the modest sensitivity and resolution of our observations prevent us from obtaining definitive constraints on the $^{12}$CO extent and dynamics of AzTEC/C159. \\
	
	\item The $^{12}$CO(5$-$4)/CO(2$-$1) line brightness temperature ratio of $r_{52}= 0.55\pm 0.15$ is consistent with  the high gas excitation of the star-forming disk galaxies GN20 \citep{carilli10, hodge12} and J1000$+$0234 \citep[][J17]{schinnerer08} at $z>4$. 
	In general, the  $^{12}$CO SLED of  AzTEC/C159 is similar to the median for SMGs \citep{bothwell13} 
and significantly  deviates from  that for star-forming disk galaxies at $z\sim1.5$ \citep{daddi15}. \\

	\item The CO$\to$\,H$_2$ conversion factor ($\alpha_{\text{CO}}$) for AzTEC/C159   is \hbox{$3.9^{+2.7}_{-1.3}\,{\rm M}_{\odot}$\,K$^{-1}$\,km$^{-1}$\,s\,pc$^{-2}$} and
	 \hbox{$3.8^{+2.6}_{-1.4}\,{\rm M}_{\odot}$\,K$^{-1}$\,km$^{-1}$\,s\,pc$^{-2}$}, as given by the dynamical and gas-to-dust method, respectively. 
	This suggests that the conditions of the ISM in AzTEC/C159 would be consistent with  those of local star-forming disks.

\end{enumerate}

The intense star formation activity of AzTEC/C159 does not seem to be triggered by major mergers as other SMGs. Instead, 
its molecular gas conditions  suggest that cold gas streams  may fuel a gravitationally unstable  gas-rich disk,  that harbors extreme SFE and  high gas excitation. A population of high-redshift disk galaxies has been predicted by both numerical simulations and recent observational studies of quiescent  rotating disk galaxies at $z\sim2$, that are believed to be the descendants of AzTEC/C159-like galaxies. \\

\begin{acknowledgements}
We thank  Ian Smail for his detailed comments and suggestions that helped to improve the manuscript. 
E.F.J.A  would like to thank M. Krips for her support during the visit to IRAM/Grenoble and the hospitality of the DSOC in Socorro/New Mexico. Support for B.M. was provided by the DFG priority program 1573 ``The physics of the interstellar medium''. G.C.J. is grateful for support from NRAO through the Grote Reber Doctoral Fellowship Program. E.F.J.A,  B.M., A.K., E.R.D. and F.B.  acknowledge  support of the Collaborative Research Center 956, subproject A1 and C4, funded by the Deutsche Forschungsgemeinschaft (DFG). C.G.G and S.T. acknowledge support from the European Research Council (ERC) Consolidator Grant funding scheme (project ConTExt, grant number: 648179).  D.R. acknowledges support from the National Science Foundation under
grant number AST-1614213.  M.A. acknowledges partial support from FONDECYT through grant 1140099. M.J.M.~acknowledges the support of  the National Science Centre,
Poland through the POLONEZ grant 2015/19/P/ST9/04010;
this project has received funding from the European Union's Horizon
2020 research and innovation programme under the Marie
Sk{\l}odowska-Curie grant agreement No. 665778. V.S. acknowledges support from the European Union's Seventh Frame-work program under grant agreement 337595 (ERC Starting Grant, ``CoSMass'').
 The National Radio Astronomy Observatory (NRAO) is operated by Associated Universities, Inc., under cooperative agreement with the National Science Foundation.  IRAM is supported by INSU/CNRS (France), MPG (Germany) and IGN (Spain). 
\end{acknowledgements}


\bibliographystyle{aa} 
\bibliography{jimenez-andrade+_aa.bib} 

 \object{AzTEC/C159}
\end{document}